\documentclass[manuscript]{emulateapj}
\usepackage{graphicx}
\usepackage{epsf}
\usepackage{amsmath, amsthm, amssymb} 

\shortauthors{Skelton, Bell, \& Somerville}
\shorttitle{Dry merging and the color--magnitude relation}

\newcommand{\msun}{${\rm M_{\sun}}$}

\newcommand{\captionfonts}{\footnotesize}

\makeatletter  
\long\def\@makecaption#1#2{%
  \vskip\abovecaptionskip
  \sbox\@tempboxa{{\captionfonts #1: #2}}%
  \ifdim \wd\@tempboxa >\hsize
    {\captionfonts #1: #2\par}
  \else
    \hbox to\hsize{\hfil\box\@tempboxa\hfil}%
  \fi
  \vskip\belowcaptionskip}
\makeatother   

\begin{document}

\title{The effect of dry mergers on the color--magnitude relation of early-type galaxies}

\author{Rosalind E. Skelton$^1$, Eric F.\ Bell$^1$ and Rachel S. Somerville$^{1,2}$}
\affil{$^1$Max-Planck-Institut f\"ur Astronomie,
K\"onigstuhl 17, D-69117 Heidelberg, Germany; skelton@mpia.de\\$^2$Space Telescope Science Institute, 3700 San Martin Drive, Baltimore, MD 21218, USA}

\begin{abstract}
We investigate the effect of dry merging on the color--magnitude relation (CMR) of galaxies and find that the amount of merging predicted by a hierarchical model results in a red sequence that compares well with the observed low-redshift relation. A sample of $\sim$29, 000 early-type galaxies selected from the Sloan Digital Sky Survey Data Release 6 shows that the bright end of the CMR has a shallower slope and smaller scatter than the faint end. This magnitude dependence is predicted by a simple toy model in which gas-rich mergers move galaxies onto a ``creation red sequence'' (CRS) by quenching their star formation, and subsequent mergers between red, gas-poor galaxies (so-called ``dry'' mergers) move galaxies along the relation. We use galaxy merger trees from a semianalytic model of galaxy formation to test the amplitude of this effect and find a change in slope at the bright end that brackets the observations, using gas fraction thresholds of 10 -- 30\% to separate wet and dry mergers. A more realistic model that includes scatter in the CRS shows that dry merging decreases the scatter at the bright end. Contrary to previous claims, the small scatter in the
observed CMR thus cannot be used to constrain the amount of dry merging.
\end{abstract}


\section{Introduction}

\par

In this Letter we explore the formation of massive red galaxies, following the effects of dry mergers (mergers between two gas-poor progenitors) on the colors and magnitudes of the galaxy population. 

\par
Galaxies are found to occupy two distinct regions in color--magnitude space, known as the red sequence and blue cloud \citep{Strateva01, Blanton03}. The blue cloud, made up mostly of star-forming late-type galaxies, is a broad distribution with large scatter in color at all magnitudes. The red sequence is made up mostly of early-type galaxies with little continuing star formation. These galaxies lie along a tight color--magnitude relation (CMR), which results primarily from the relation between mass and metallicity \citep[e.g.,][]{Faber73, Larson74, Kodama97, Gallazzi06}, in the sense that the most massive galaxies are the most metal rich and consequently redder. 
\par
The amount of stellar mass in the red galaxy population has approximately doubled since $z=1$ \citep[e.g.,][]{Bell04, Brown07, Faber07}; yet these galaxies have low levels of star formation which cannot account for the increase in mass. In contrast, the amount of stellar mass in blue galaxies remains approximately constant over the same time period, although these galaxies are actively forming stars. Much of the growth of the red sequence population can be accounted for by the truncation of star formation in $<L^*$ galaxies \citep{Bell07}. A fairly rapid process is needed to transform both the colors and morphologies of these galaxies, moving them from the blue cloud onto the red sequence. One such mechanism is the merging of galaxies, a natural consequence of hierarchical structure growth. This results in remnants that are reddened through the loss of gas and subsequent slowing of star formation and have randomized orbital motions \citep[e.g.,][]{Toomre72, Barnes96, Cox06}. 

In contrast, the most massive $>L^*$ galaxies on the CMR are thought to form through mergers of galaxies that already lie on the red sequence and contain little gas. Models of galaxy formation predict these dry mergers to play an important role and they have been observed from their morphological signatures \citep{vanDokkum05, Bell06a, McIntosh08} but concensus has not yet been reached on the observed merger rate and resulting growth of mass since $z\sim1$. Merger rates based on close-pair counts indicate significant merger activity between $\sim L^*$ early-type galaxies \citep[e.g.,][]{vanDokkum05, Bell06a, McIntosh08} with major dry mergers playing an increasing role towards lower redshifts and for more massive galaxies \citep[e.g.,][]{Lin08, Bundy09}. A key uncertainty in this approach is the assignment of a merging timescale; this uncertainty is considerable, exceeding a factor of 2. Other observations can indirectly constrain the dry merger rate; e.g., the evolution of the stellar mass--size relation \citep{McIntosh05, vanderWel08} and number density of the most massive early-type galaxies since $z\sim1$ \citep[e.g.,][]{Scarlata07, Cimatti06, Wake06, Cool08, Faber07}. 

The CMR provides another promising avenue to explore. Both the slope and scatter of the relation place constraints on the formation histories of early-type galaxies. The CMR is generally assumed to be linear; however, there is some debate whether there is a change of slope with magnitude. This is evident in the CMRs of some clusters \citep[e.g.,][]{Metcalfe94, Ferrarese06} but a number of other determinations show no particular evidence for a break with magnitude \citep[e.g.,][]{Terlevich01, McIntosh05}. In the field, \citet{Baldry04} described the low-redshift red sequence of Sloan Digital Sky Survey \citep[SDSS;][]{York00} galaxies by a tanh function plus a straight line. The scatter in the relation ranges from as little as 0.04 mag for the Coma and Virgo clusters \citep{Bower92, Terlevich01} to 0.1 mag in other clusters and the field \citep{Schweizer92, McIntosh05, Ruhland09}. The intrinsic scatter limits the spread in age of the stellar populations \citep[][BKT98 hereafter]{Bower92, BKT98}. This was seen as evidence that elliptical galaxies formed at high redshifts, evolving passively thereafter; however, recent work shows that this scatter is consistent with a model for the constant growth of the red sequence through the quenching of star formation in blue cloud galaxies \citep{Harker06, Ruhland09}. BKT98 used a simple model to argue that dry merging would cause a decrease in slope and increase in the scatter of the relation. The tightness of the relation in clusters such as Coma would thus limit the amount of mass growth due to dry mergers to a factor of 2--3 at most since the majority of stars were formed.

\par
We reconsider the effect of dry mergers on the bright end of the CMR. As noted by \citet{Bernardi07a}, color is not expected to change during such mergers since there is no associated star formation; thus galaxies move horizontally as the mass of the system increases. In addition, the shape of the mass function suggests that a bright galaxy will preferentially merge with one of the more numerous fainter galaxies. These galaxies lie further down the CMR and are bluer; thus dry mergers cause a tilt toward bluer colors at the bright end. We show that this is indeed observed in the local CMR averaged over all environments, using data from the SDSS (Section 2). In the same spirit as BKT98, we use a simple toy model to investigate the consequences of dry merging, using galaxy merger histories from a semianalytic model (SAM) of galaxy formation in a $\Lambda$CDM hierarchical universe (Section 3). We show that when scatter is included in the initial CMR, dry merging causes a decrease in the slope and a tightening of the CMR at the bright end. Although dry merging does affect the evolution of the CMR, the predicted effect does not conflict with the observed relation. We adopt a cosmology with $\Omega_M=0.3$, $\Omega_\Lambda=0.7$, and $H_0=100h$~km~s$^{-1}$~Mpc$^{-1}$ with $h=0.7$ throughout.

\section{The observed red sequence}\label{obs.sec}

We use a subsample of galaxies from the SDSS Data Release 6 \citep[DR6;][]{Adelman-McCarthy08} selected from the New York University Value-Added Galaxy Catalog \citep[NYU-VAGC;][]{Blanton05b}. We select galaxies in a thin redshift slice ($0.0375<z<0.0625$) with Galactic extinction corrected \citep{Schlegel98} Petrosian magnitudes $m_r<17.77$, resulting in a sample of 72,646 objects. This range in redshift provides a significant number of bright galaxies but is narrow enough to avoid the need for volume and evolution corrections. We use the S\'ersic magnitudes provided in the NYU-VAGC as an estimate of total magnitude, since Petrosian magnitudes are known to underestimate the total flux, particularly for early-type galaxies \citep{Graham05}. The photometry of bright galaxies in the SDSS is also affected by the overestimation of the sky background, which leads to underestimated magnitudes \citep[see, e.g.,][]{Bernardi07a}. The change in slope seen at the bright end of the CMR would be enhanced by a systematic brightening of the magnitudes of the most massive galaxies; thus we expect our result to be robust to these photometric errors. We $k$-correct to rest-frame $z=0.1$ bandpasses using \texttt{kcorrect\_v4\_1} \citep{BlaRow07}. The most reliable estimate of galaxy color uses SDSS model magnitudes, where the $r$-band image is used to determine the best-fit (exponential or de Vaucoleurs) profile and only the amplitude of the fit adjusted in other bands. An equivalent aperture is thus used in both bands, resulting in an unbiased and high S/N estimator of color in the absence of gradients across the galaxy. All methods using the same aperture in different bands (fiber magnitudes, model magnitudes or S\'ersic magnitudes defined using the i-band S\'ersic model as a convolution kernel for the other bands), as well as Petrosian magnitudes, give a very similar result. S\'ersic models fit to each band separately give much greater scatter, and qualitatively similar curvature, compared to the higher S/N model magnitudes we use here.

We have chosen to show the CMD in $g-r$ color and $r$-band magnitude space because both the bimodality and a change in slope are obvious features; however, we verified that these are also visible in other combinations of color and magnitude. Concentration, given by $C=R_{90}/R_{50}$ where $R_{90}$ and $R_{50}$ are the radii enclosing 90\% and 50\% of the Petrosian flux, respectively, has been shown to correlate with galaxy morphology \citep{Strateva01, Shimasaku01}. We apply a cut of $C\ge2.6$, a criterion that has been used to broadly select early-type galaxies in previous work using SDSS data \citep[e.g.,][]{Bell03, Kauffmann03a}, to isolate the red sequence. This ensures that a Gaussian fit to the CMR is not pulled down by late-type galaxies at the faint end, where the separation between the two populations is less distinct. The remaining sample contains 29,017 galaxies and is complete for $M_r\la-18.3$~mag. 

The CMD for these galaxies is shown in the upper left panel of Figure~\ref{cmd.fig}. In each magnitude bin of 0.25 mag along the red sequence, we fit a Gaussian function\footnote{We have verified that fitting a double Gaussian to account for residual blue cloud galaxies does not affect the position of the mean.} to the distribution of colors. The mean and width of the Gaussians are shown as diamonds and bars in the figure.  It is clear that the slope of the relation changes with magnitude, flattening at the bright end. We fit straight lines to the means above and below $M_r=-21$~mag and show these as dashed lines in all panels to facilitate comparison with the model. The faint-end fit, $^{0.1}(g-r)=0.10-0.04^{0.1}M_r$, is used as the input creation red sequence (CRS) for the model, as described below. 

\section{Modeling the effect of merging along the red sequence}

We would like to isolate the effect of dry merging on the colors of galaxies on the red sequence through a simple toy model. This approach is similar to previous work by BKT98; however, we use galaxy merger histories from an up-to-date model of galaxy formation and make different assumptions about the formation of the red sequence. We suppose that quenching of star formation -- either through merging between gas-rich galaxies (``wet'' mergers) or dwindling gas supply -- places galaxies on a CRS, whereas the BKT98 model assumes that the relation is in place at some formation epoch. Galaxy merger trees and information on the masses and gas fractions of merging galaxies are extracted from the \citet{Somerville08} SAM. The full SAM, which incorporates star formation and feedback effects, produces a bimodal color distribution in broad qualitative agreement with observations. We make simple assumptions to determine the magnitudes and colors of galaxies on the red sequence after merging events rather than using the luminosities given by the full model, however. This enables us to attribute the effects directly to merging rather than trying to disentangle the complex mix of baryonic processes included in the SAM.

\begin{figure}
\plotone{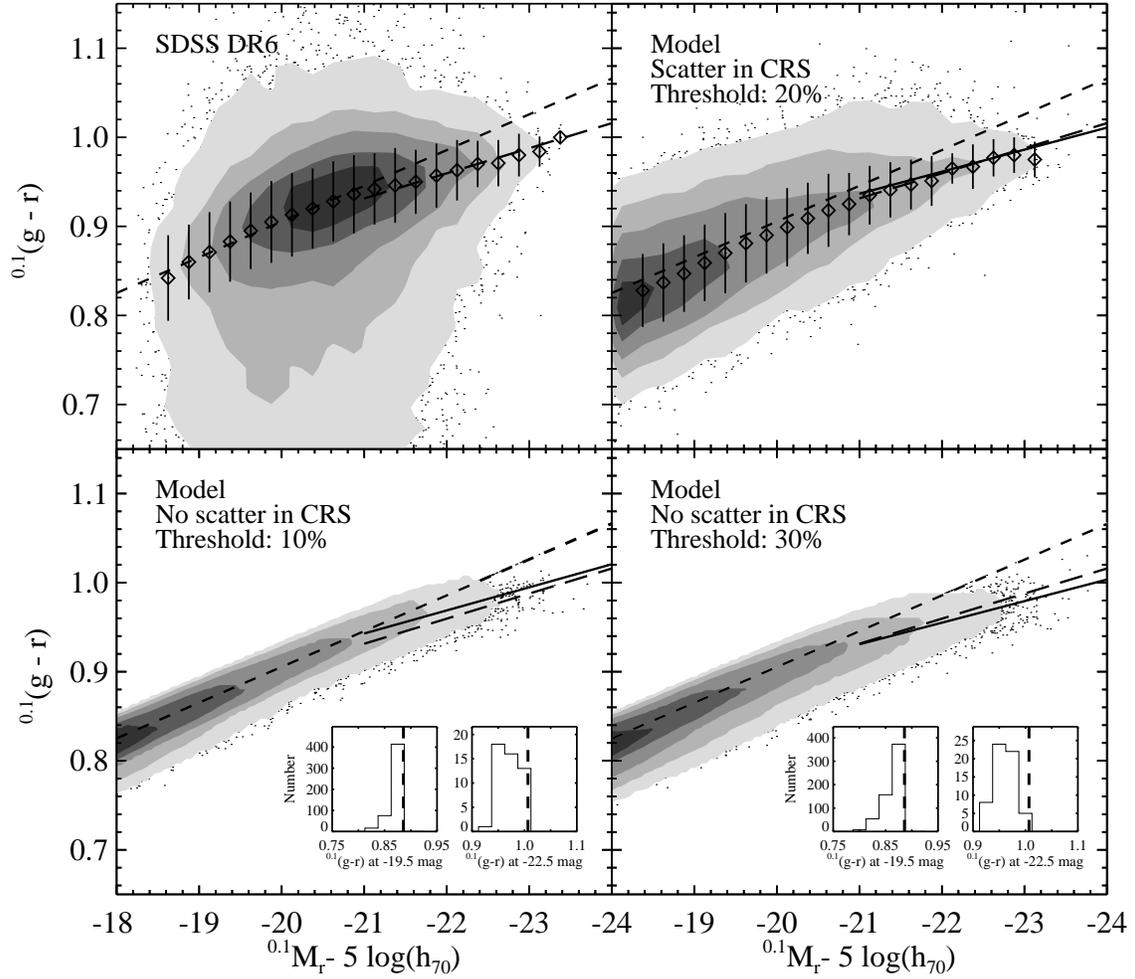}
\caption{Red sequence from SDSS observations compared to a toy model for dry merging. The upper left panel shows the CMD of galaxies with concentrations of $C\geq2.6$ from the SDSS DR6. The upper right panel shows the red sequence for a model which includes scatter in the CRS, with a gas fraction threshold of 20\%. The contours enclose 2, 10, 25, 50 and 75\% of the maximum value. The mean and scatter of the distribution binned in magnitude are shown as diamonds and bars in the upper panels. The short dashed lines in all panels show the fit to the observed means for magnitudes fainter than $M_r=-21$~mag, extended over the whole magnitude range to illustrate the change in slope at the bright end. Long dashed lines show the fit to the bright end ($M_r<-21$~mag) of the observed relation while fits to the model distributions are shown as solid lines. The lower panels show the model red sequence using gas fraction thresholds of 10 and 30\% with no scatter in the initial relation. The inset histograms show the distribution of colors in two magnitude bins 0.1 mag wide, centered on $^{0.1}M_r=-19.5$ and $^{0.1}M_r=-22.5$ mag. In the faint bin, most galaxies still lie on the CRS (dashed lines), while in the bright bin most galaxies have undergone dry mergers and moved off the initial relation.\label{cmd.fig}} 
\end{figure}

Dark matter (DM) merger histories are constructed using the extended Press--Schechter formalism, as described in \citet{SK99} but with the modifications described in \citet{Somerville08}, which lead to better agreement with N-body simulations. Galaxies within the merged DM haloes lose angular momentum through dynamical friction and fall toward the center, merging some time later \citep[see][for details]{Somerville08}. We follow the merger histories of galaxies within all haloes with ${\rm{log}_{10}[M_{halo}}/$\msun]$>11.7$, rather than just cluster-sized haloes as BKT98 did. We consider mergers with mass ratios between 1:1 and 1:10, where the mass used is the DM mass within twice the characteristic NFW scale radius plus the total baryonic mass; however, only major mergers (mass ratios between 1:1 and 1:4) are assumed to be sufficient at quenching star formation resulting in a remnant on the CRS. The magnitude of the remnant galaxy is found from the total stellar mass of the two progenitors using the M/L ratio of low-redshift red sequence galaxies produced by the SAM ($^{0.1}M_r=2.87-2.22 {\rm{log}_{10}[M_*}/$\msun]). For each of the progenitor galaxies, we determine the fraction of baryonic mass contained in cold gas. If either galaxy has a gas fraction above some threshold, the merger is assumed to be wet and produces a remnant galaxy on the CRS.  
In order to compare directly with observations we have used the measured faint-end slope and zeropoint of the observed red sequence to specify the remnant's color. The CRS is thus given by $^{0.1}(g-r)=0.10-0.04^{0.1}M_r$, as described in Section~\ref{obs.sec} and shown by the short dashed lines in all panels of Figure~\ref{cmd.fig}. In so choosing, we assume that all gas-poor galaxies appear on the CRS at the epoch of interest, and evolve passively (and identically) thereafter.  Accordingly, one can parameterize a galaxy completely by the $z=0.1$ color and magnitude. Subsequent dry merging, where the gas fraction is below the chosen threshold, produces remnant galaxies with colors and magnitudes determined by the simple combination of the progenitor colors and magnitudes. In cases where a galaxy's recorded merger history has no major wet mergers, the first gas-poor progenitors to merge are assigned colors on the CRS and the remnant's properties are determined as above. 

In Figure~\ref{cmd.fig} we compare the observed red sequence (upper left panel) with the CMR of remnant galaxies in the model, using gas fraction thresholds ranging from 10 to 30\%. A linear regression to the bright end ($M_r < -21$ mag) of the model distribution is shown as a solid line in each case. There is a clear tilt toward a shallower slope for bright galaxies produced by merging, with the slope and break point sensitive to the chosen gas fraction threshold. The observed bright-end slope is bracketed by models with thresholds of 10\% and 30\% (lower panels). Using a lower threshold, more mergers are defined as wet, with remnants placed directly onto the CRS, and thus there are fewer remnants of dry mergers with bluer colors than the initial relation. This can also be seen in the inset histograms in the lower panels, which show the distribution of colors in magnitude bins of 0.1 mag centered at $^{0.1}M_r=-19.5$ and $^{0.1}M_r=-22.5$ mag. In the faint bin, the distribution peaks on the CRS (dashed line) showing that most of the faint galaxies have not had dry mergers. In contrast, most of the galaxies in the bright bin have undergone dry mergers and are predicted to lie significantly blueward of the CRS at the present day. The shift away from the CRS is smaller when the gas fraction threshold is lower; furthermore, the slope at the bright end changes less dramatically and the break from the CRS occurs at brighter magnitudes. 

The upper right panel of Figure~\ref{cmd.fig} shows the effect of including scatter in the initial relation, with a gas fraction threshold of 20\%. We assume that galaxies produced by wet mergers are normally distributed about the CRS, with the width of the distribution given by the average observed scatter for galaxies with $M_r>-21$ mag (0.046~mag). We fit a Gaussian to the color distribution in each magnitude bin of 0.25 mag, in the same way as for the observations. The means and widths of these Gaussians are shown as diamonds and bars in the figure. The slope at the bright end decreases as expected from the models without scatter, while the relation becomes tighter, with decreasing scatter toward the bright end. The means of the Gaussians are offset slightly from the observed relation due to the small fraction of dry mergers taking place at all magnitudes. Applying a small zeropoint offset ($\sim0.01$ mag) to the CRS with a corresponding change in gas fraction threshold would account for this difference without affecting the result. The scatter around the mean in the most luminous bins is slightly smaller in the observations; however, we note that there are very few galaxies in the brightest bins and thus the Gaussian fits are more tentative. 

\section{Discussion and conclusions}

The existence of a tight CMR over a wide range in magnitude has been used to argue against the importance of dry mergers for the growth of red galaxies, since they were expected to flatten the relation and increase the scatter (BKT98); yet there seem to be few plausible alternative mechanisms for the creation of the most massive galaxies on the red sequence. We revisit this apparent conflict, showing that the red sequence for local field galaxies from the SDSS has a change in slope at the bright end and that a toy model for dry merging in a hierarchical universe produces a red sequence that is consistent with these observations. 

In our simplest model, remnants of wet mergers are placed on a CRS with no scatter. Subsequent dry merging results in a tilt toward bluer colors at the bright end and an increase in the width of the relation as remnants move off the initial line. A change in the slope of the CMR was predicted by the model of BKT98; they assumed that the initial red sequence formed at a given {\it time} and that subsequent merging at all stellar masses was dry, shifting the entire population to bluer colors. In contrast, we find that the relation will flatten only at the bright end. We associate the formation of the red sequence with an {\it event} (a gas-rich merger). This leads to little change at the faint end (most mergers there are gas-rich) and a stronger change at the bright end (most mergers there are dry). The change in slope and magnitude at which the break occurs depend strongly on the assumption of a gas fraction threshold below which mergers are assumed to be dry. We find that gas fraction thresholds of 10\% and 30\% bracket the observed relation. The brightest galaxies in all environments (halo masses) experience growth through dry mergers thus we do not find an offset for brightest cluster galaxies compared to the rest of the population at the same luminosity \citep[cf.][]{Bernardi07a}.

A model including scatter in the CRS shows a similar change in slope and a reduction in scatter at the bright end as a consequence of the central limit theorem. The width of the relation is thus only increased by dry merging when the initial relation is assumed to have no scatter. The scatter in the observed relation is slightly smaller than in the model; however, we have not accounted for differences in the age or metallicity of the galaxies involved in mergers. We have assumed the CRS has the same scatter as the faint end of the observed red sequence of local galaxies, which has a contribution from the aging of the stellar populations \citep{Gallazzi06}. If dry mergers occur soon after galaxies arrive on the red sequence, the difference in color between merging galaxies will be smaller than the scatter of the total population and thus the scatter in color of dry merger remnants will be even smaller than we predict. 

\acknowledgments
We thank the referee for a useful report. R.E.S. is a member of the International Max Planck Research School for Astronomy \& Cosmic Physics at the University of Heidelberg. This publication makes use of the Sloan Digital Sky Survey (http://www.sdss.org/).

\clearpage
\end{document}